\documentclass{PoS}

\usepackage{fontenc}
\usepackage{graphicx}
\usepackage{amsmath,amsfonts,bm}
\usepackage{amssymb}
\usepackage{slashed}
\usepackage{subfigure}
\usepackage{color}
\usepackage{latexsym}

\title{Pion Electromagnetic  Form Factor in Virtuality Distribution Formalism}

\ShortTitle{Pion EM Form Factor \& Virtuality Distributions}

\author{\speaker{A. V. Radyushkin }\\
     Physics Department, Old Dominion University, Norfolk,
             VA 23529, USA \\ 
Thomas Jefferson National Accelerator Facility,
              Newport News, VA 23606, USA\\
E-mail: \email{radyush@jlab.org}}

\abstract{We discuss two applications of the {\it Virtuality Distribution Amplitudes} (VDA)
 formalism developed in our recent papers.  
We  start with an overview of the 
main properties of the 
pion distribution amplitude  emphasizing 
the quantitative measures of its width, and possibility to access them through 
the pion transition form factor studies.   We formulate the basic concepts of 
the VDA approach and  introduce the pion {\it transverse momentum distribution amplitude} 
(TMDA) which plays,  in a covariant Lagrangian formulation,   a role similar 
to that of the pion wave function in the  3-dimensional Hamiltonian light-front approach.
 We propose simple factorized models for soft TMDAs, 
 and use them to describe existing data on the pion transition form factor,
 thus fixing the scale determining the size of the  transverse-momentum effects. 
 Finally, we apply the VDA approach to the 
 one-gluon exchange 
contribution for the pion electromagnetic form factor.   
We observe a very late $Q^2 \gtrsim 20$ GeV$^2$ onset of transition to 
 the asymptotic pQCD predictions and show that in the $Q^2 \lesssim 10$ GeV$^2$
region there is essentially no sensitivity to the shape of the pion distribution amplitude. 
Furthermore, the magnitude of the one-gluon exchange 
contribution in this region is estimated to be an order of magnitude 
below the  Jefferson Lab data, thus leaving the Feynman mechanism 
as the only one relevant to    the pion electromagnetic  form factor behavior 
for accessible $Q^2$.
}

\FullConference{QCD Evolution 2015 -QCDEV2015- \\
		26-30 May 2015\\
		Jefferson Lab (JLAB), Newport News Virginia, USA}

\begin{document}

%

%





 \maketitle

 \section{Introduction}

Adding transverse momentum degrees of freedom
is a  popular subject in   the theory of 
hard reactions.   In reality, such an adding is not always necessary. 
We mean that  when the standard (``collinear") perturbative QCD (pQCD)
 factorization works,
one deals with objects like usual parton  densities 
(for inclusive processes) and  hadron distribution amplitudes (for exclusive processes)
none  of which explicitly depends on  transverse momenta. 
Explanations  ``to public''  that   these functions  result  from  integrating 
some  more general transverse momentum-dependent functions over $k_\perp$ 
may help to create an intuitive picture,
but  in fact  all the functions  entering into the collinear
factorization formulas  are defined without any reference
to  transverse momentum, 
say, through matrix  elements of bilocal operators on the light cone.
Still, it might be    helpful  to have ``underlying''  formulas, in which transverse momentum
enters explicitly.  

An important example is the pion distribution
amplitude, which is  usually \cite{Radyushkin:1977gp}  defined  through a matrix element 
of a bilocal operator on the light cone, but sometimes  \cite{Lepage:1979zb}  is introduced 
through the light-front  wave
function $\psi (x, k_\perp)$ integrated over transverse momentum $k_\perp$.
A subtle point  is that the  first  definition is made within the 
operator product expansion  (OPE) formalism  
 of the  
 covariant 4-dimensional Lagrangian quantum field theory (QFT), while the   wave 
function   $\psi (x, k_\perp)$  
 is an object  of a  very different  Hamiltonian 
 3-dimensional  light-front (LF)  approach \cite{Drell:1969km,Lepage:1980fj}.

In our recent papers \cite{Radyushkin:2014vla,Radyushkin:2014xaa,Radyushkin:2015wqa,Radyushkin:2015},
 we have  developed 
a formalism  which is  based on   a  covariant 
QFT in  4 dimensions,  but  still manages to  encode information about  the  
hadronic structure  in hard exclusive processes  in terms of  
{\it transverse momentum dependent distribution amplitudes} (TMDAs) 
 $\Psi (x, k_\perp)$  that  incorporate  the  dependence 
on the  transverse momentum of its constituents.   
Just like in the light-front formalism, the organization of these 
 functions has the structure of the Fock state decomposition,
 i.e. each function is characterized by the number of constituents 
 involved. 

For exclusive processes, a standard situation  asking  to add  transverse momentum 
degrees of freedom is 
when the collinear factorization integral  diverges at the end point, and 
hence   one needs some    natural source  of a cut-off.
In this talk, we discuss  the use of the TMDA formalism 
 in the description of the photon-pion transition  and pion electromagnetic form factors.

\section{Pion Distribution Amplitude}

Within the covariant QFT, the pion distribution amplitude (DA) $\varphi_\pi (x)$ is 
 \mbox{introduced \cite{Radyushkin:1977gp}}  as a function
  whose $x^N$ moments  $f_N(\mu)$ 
 are  given by matrix elements of  twist-2 local  operators 
\begin{align}
f_N (\mu) =  \int_0^1 x^N \, \varphi_\pi  (x,\mu) \,  dx 
\ \ , \ \ 
i^{N+1}   \left \langle 0 | \bar d (0) \gamma_5 \slashed n   (n D)^N u (0) | \pi^+, P 
\right \rangle 
=  (Pn)^{N+1} \, f_N (\mu)  \  ,
\end{align}
with $n^2=0$ and $\mu$ being the UV renormalization scale for operators. 
Since 
the zeroth moment , 
\begin{align}
  \int_0^1  \varphi_\pi  (x,\mu) \,  dx = f_\pi \ , 
\label{fpi}
\end{align}
  is   given by the pion decay constant $f_\pi$,
we have  an important constraint
on the pion DA, fixing the integral under the
$ \varphi_\pi  (x)$ curve,  but it puts no restrictions 
on its shape. In fact, the pion DA depends on    
the renormalization scale $\mu$: $  \varphi_\pi  (x) \to \varphi_\pi  (x, \mu)$.
The solution  of the evolution equation for the pion DA  
 was obtained \cite{Efremov:1979qk,Lepage:1979zb} 
in the form of   expansion over Gegenbauer polynomials
\begin{align}
 \varphi_\pi  (x,\mu) = 6 f_\pi \, x (1-x) \,  \left \{ 1+ \sum_{n=1}^\infty a_{2n} C_{2n}^{3/2} (2x-1) 
\biggl [ \ln (\mu^2/\Lambda^2) \biggr ]^{-\gamma_{2n} / \beta_0} \right \} \  . 
\label{Gegen}
\end{align}
Since  $\gamma_{2n}>0$,    the pion DA acquires a simple form \cite{Efremov:1978fi}
$
 \varphi_\pi  (x, \mu\to \infty) = 6 f_\pi \, x (1-x) \, 
$ 
(known as the ``asymptotic DA'')  when the normalization scale
$\mu$  tends to infinity.

A   quantitative measure  of the width of the pion DA  at low normalization scales \mbox{$\mu \lesssim 1$ \,GeV }  is given 
by its  moments $\langle \xi^2 \rangle $ and  $\langle \xi^4 \rangle $  in the relative variable $\xi \equiv x -(1-x)$. Namely, \\
$\langle \xi^2 \rangle =0 $ for infinitely narrow DA 
$\varphi^{\rm narrow}_\pi (x) = f_\pi \delta (x-1/2)$; \\ 
 $\langle \xi^2 \rangle =1/5$  for  asymptotic DA 
$\varphi_\pi (x) = 6 f_\pi x (1-x) $; \\ 
$\langle \xi^2 \rangle =1/4$  for  ``root''  DA 
$\varphi_\pi (x) = \frac{8}{\pi} f_\pi\sqrt{ x (1-x)} $; and  \\ 
$\langle \xi^2 \rangle = 1/3$  for flat DA $\varphi_\pi (x) = f_\pi$. 

\begin{figure}
\centerline{\includegraphics[width=2in]{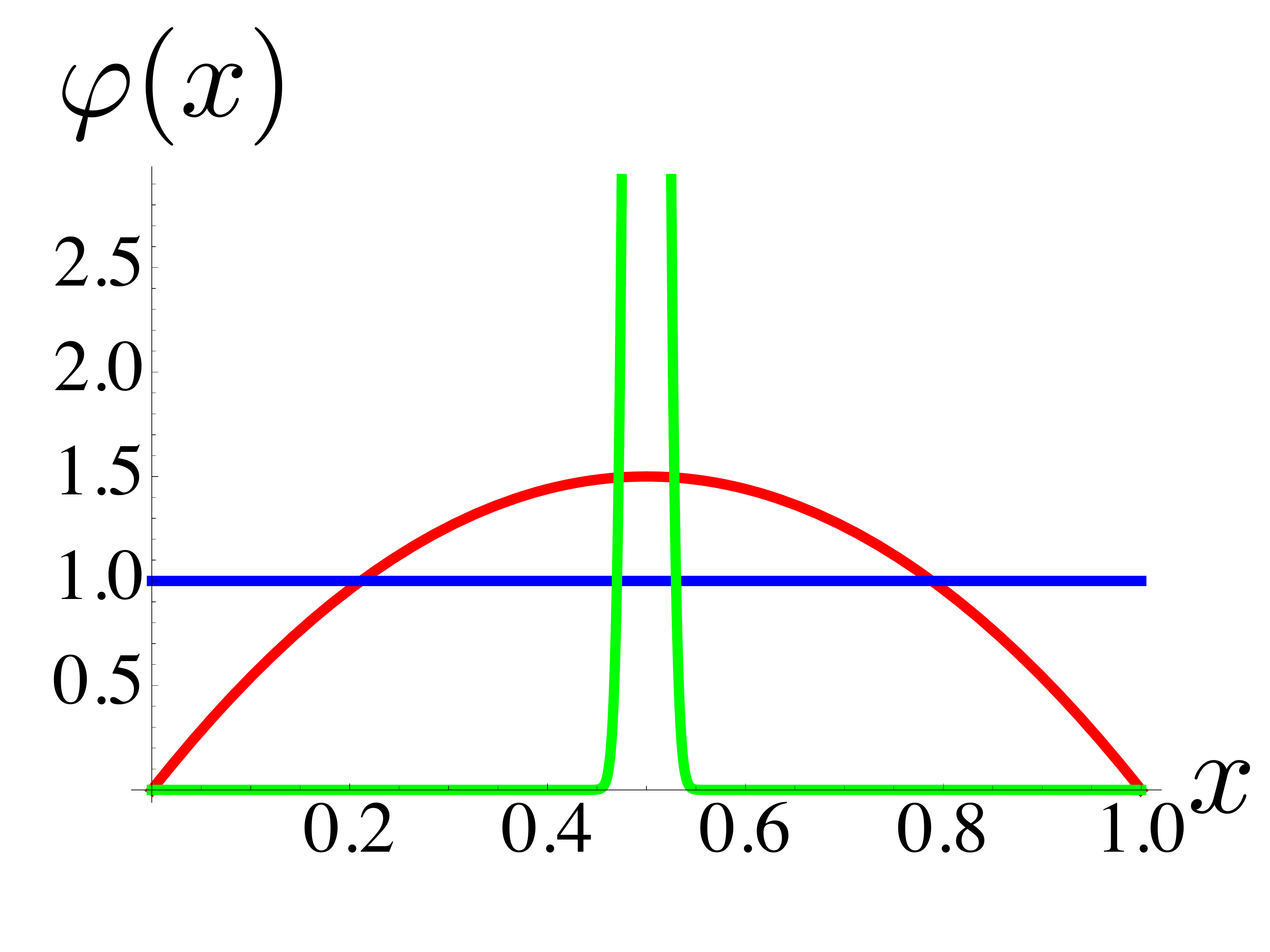} \hspace{1cm} \includegraphics[width=2in]{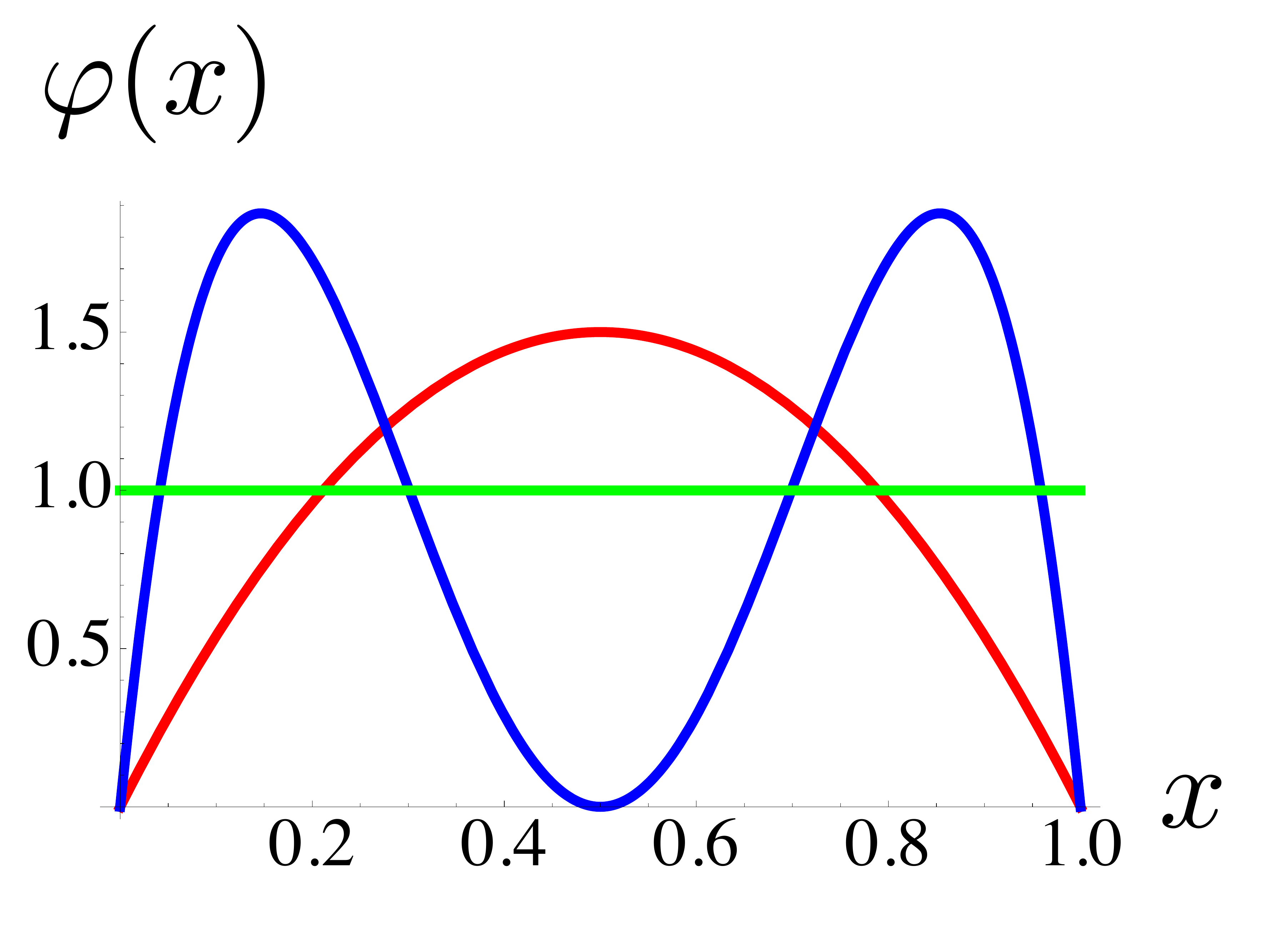}}
\caption{Shapes for pion distribution amplitude. On the left: narrow (green), asymptotic (red), flat (blue).
On the right:  asymptotic (red), flat (green),  Chernyak-Zhitnitsky (blue).}
\label{pionda} 
\end{figure}

In particular, a    QCD sum rule calculation of Chernyak and Zhitnitsky (CZ)  
gave the result 
$\langle \xi^2 \rangle=0.40 $
(at    $\mu^2 = 1.5$\,GeV$^2$),  which is larger than the value 
$1/3$ of flat DA. 
The fitting model 
$
 \varphi_\pi^{\rm CZ} (x) =30 f_\pi x (1-x) (1-2 x)^2  
$
 was constructed from the sum of two first 
terms of the Gegenbauer expansion (\ref{Gegen}), 
which has $x(1-x)$  as an overall factor,
thus excluding all models with DA's that do not linearly vanish
at the  end-points. However, 
there is no {\it a priori} principle justifying  such a  requirement:  it 
is just an assumption which  may or may not be true.

One can rewrite QCD sum rule   for  DA itself rather than for its moments:
 \begin{align}
& f_\pi\varphi_\pi(x)=\frac{3M^2}{2\pi^2}(1-e^{-s_0/M^2})x(1-x)
		 +\frac{8}{81}\frac{\pi\alpha_s\langle\bar
		   qq\rangle^2}{M^4}
\Bigl \{11 \delta(x)
		 +2\delta^\prime(x)+ \{ x \to 1-x \} \Bigr \} \ , 
		 \nonumber 
\end{align}
demonstrating that the widening of the pion DA in CZ calculation is produced by 
delta-function terms due to the quark and gluon 
condensates taken in local approximation. Bringing in the 
 non-locality of  condensates changes $\delta (x) \to{2 x}\theta (x < \Delta) /{\Delta^2}  $
with 
$ \Delta = \lambda_q^2/2 M^2$ and $\lambda_q^2$ fixed from  the 
ratio ${\lambda_q^2}  = \langle \bar q D^2 q \rangle / \langle \bar q  q \rangle
\approx 0.4$ GeV$^2$.  This modification \cite{Mikhailov:1986be, Mikhailov:1991pt}
decreases resulting $ \langle \xi^2 \rangle$ to \mbox{$\approx 0.25 $. }
This value (obtained almost 30 years ago)  is in a complete  agreement  with the 
most recent lattice result  \cite{Braun:2015axa}:  \mbox{$ \langle \xi^2 \rangle=0.24$}
  at $\mu = 2$ GeV.  
The  ``root'' model  
 \begin{align}
 \varphi_\pi^{\rm root} (x) = \frac{8}{\pi} \, f_\pi  \sqrt{x (1-x)} 
\end{align}
producing $ \langle \xi^2 \rangle=\frac14$  was proposed in our 1986 paper   \cite{Mikhailov:1986be}.
The magnitude of this  DA in the middle $
 \varphi_\pi^{\rm root} (1/2) /f_\pi = {4}/{\pi} \approx 1.27$  is close to the $ \approx 1.2$  value  found by    Braun and  Filyanov 
 in 1988 \cite{Braun:1988qv}.

\section{Photon-pion  transition form factor}

There  is also  a hope that  information  about the shape of  pion DA
can   be  extracted from experimental data,  in  particular from the 
 form factor $F_{\gamma^* \gamma^*  \pi^0}(q_1^2,q_2^2)$. 
 \begin{figure}[ht]
\centerline{\includegraphics[width=2in]{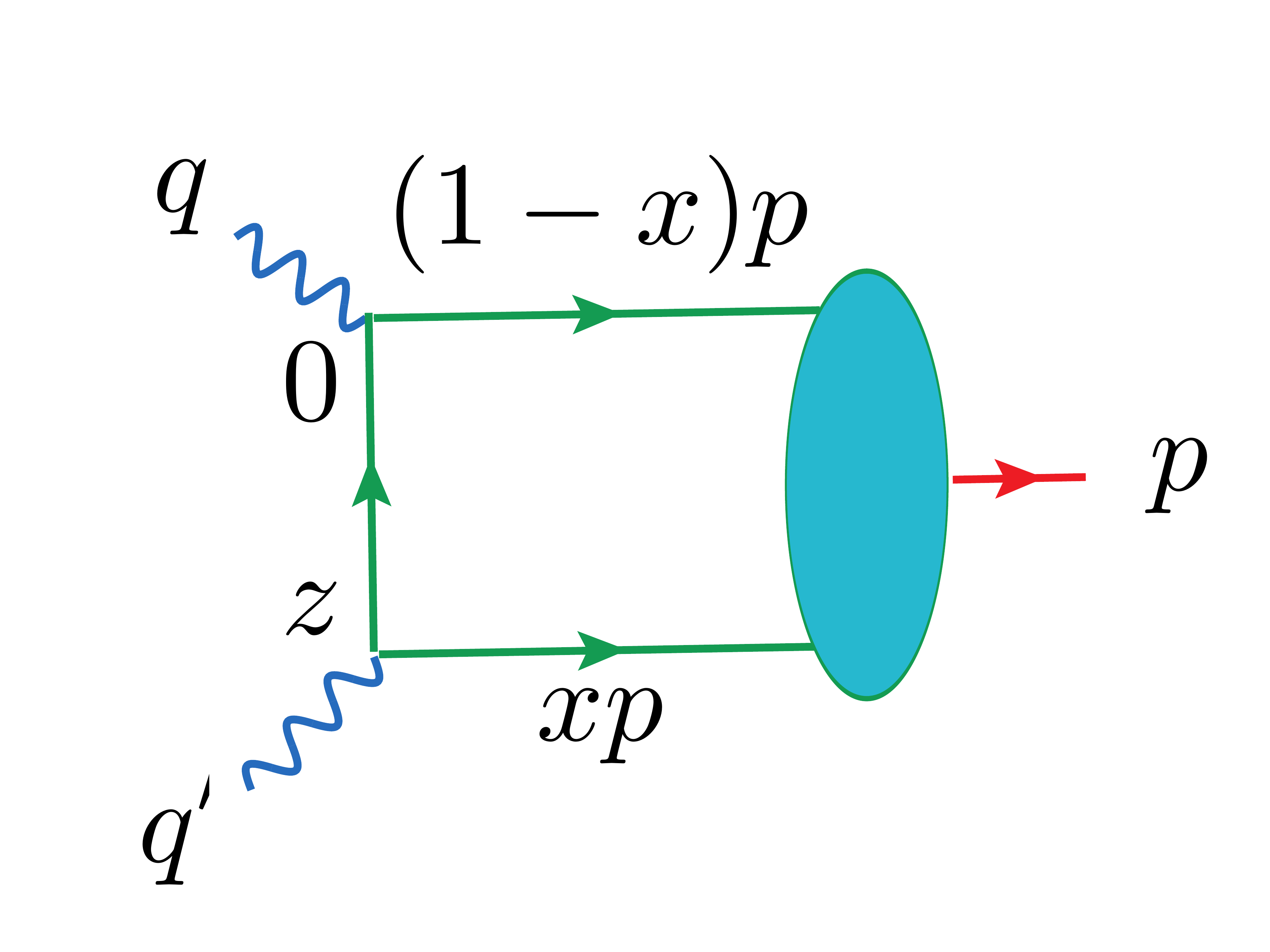}}
\caption{Parton  picture for handbag diagram describing  photon-pion transition form factor.
}
\label{pigg}
\end{figure}
If one of  the photons is  real, the leading-order perturbative QCD (pQCD)
prediction \cite{Lepage:1980fj}  is 
\begin{equation}  
F_{\gamma^* \gamma \pi}^{\rm pQCD} (Q^2) = 
\frac{\sqrt{2}}{3Q^2} \int_0^1 
 \frac{\varphi_{\pi}(x)}{x} \, dx \  \equiv  \frac{\sqrt{2}f_\pi}{3\, Q^2}\, J \ .
 \label{8b} 
\end{equation}
Information about the shape of the pion DA is  now accumulated in the factor $J$.
Namely, 
 \\ 
$J^{\rm narrow} =2$  for the infinitely narrow DA,  $\varphi^{\rm narrow}  (x) = f_\pi  \delta (x-1/2)$; \\
$J^{\rm as}=3$ 
for asymptotic DA  $\varphi^{\rm as} (x)= 6 f_\pi x (1-x)$, \\
$J^{\rm root} =4$ for   ``root''  DA 
$\varphi_\pi (x) = \frac{8}{\pi} f_\pi\sqrt{ x (1-x)} $, while  \\ 
$J^{\rm CZ}=5$ for  the CZ model  $\varphi^{\rm CZ} (x)= 30 f_\pi x (1-x) (1-2x)^2$.\\
Experimentally, 
 $F_{\gamma^* \gamma^*  \pi^0}(q_1^2\approx 0, q_2^2=-Q^2)$ 
 was measured recently by {\sc BaBar} \cite{Aubert:2009mc} 
and Belle \cite{Uehara:2012ag} collaborations.
\begin{figure}[b]
\centerline{\includegraphics[width=1.5in]{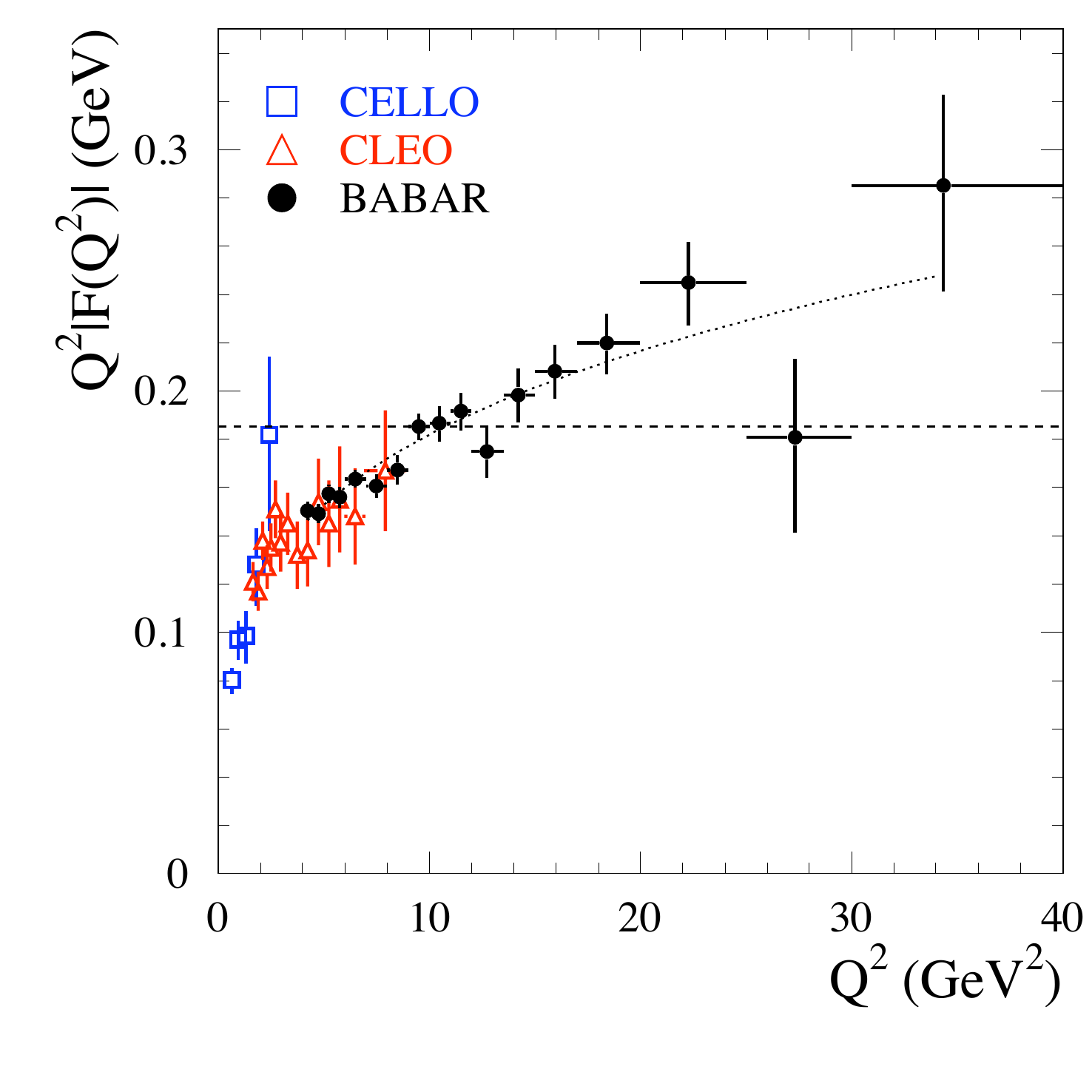}   \hspace{2cm} \includegraphics[width=2in]{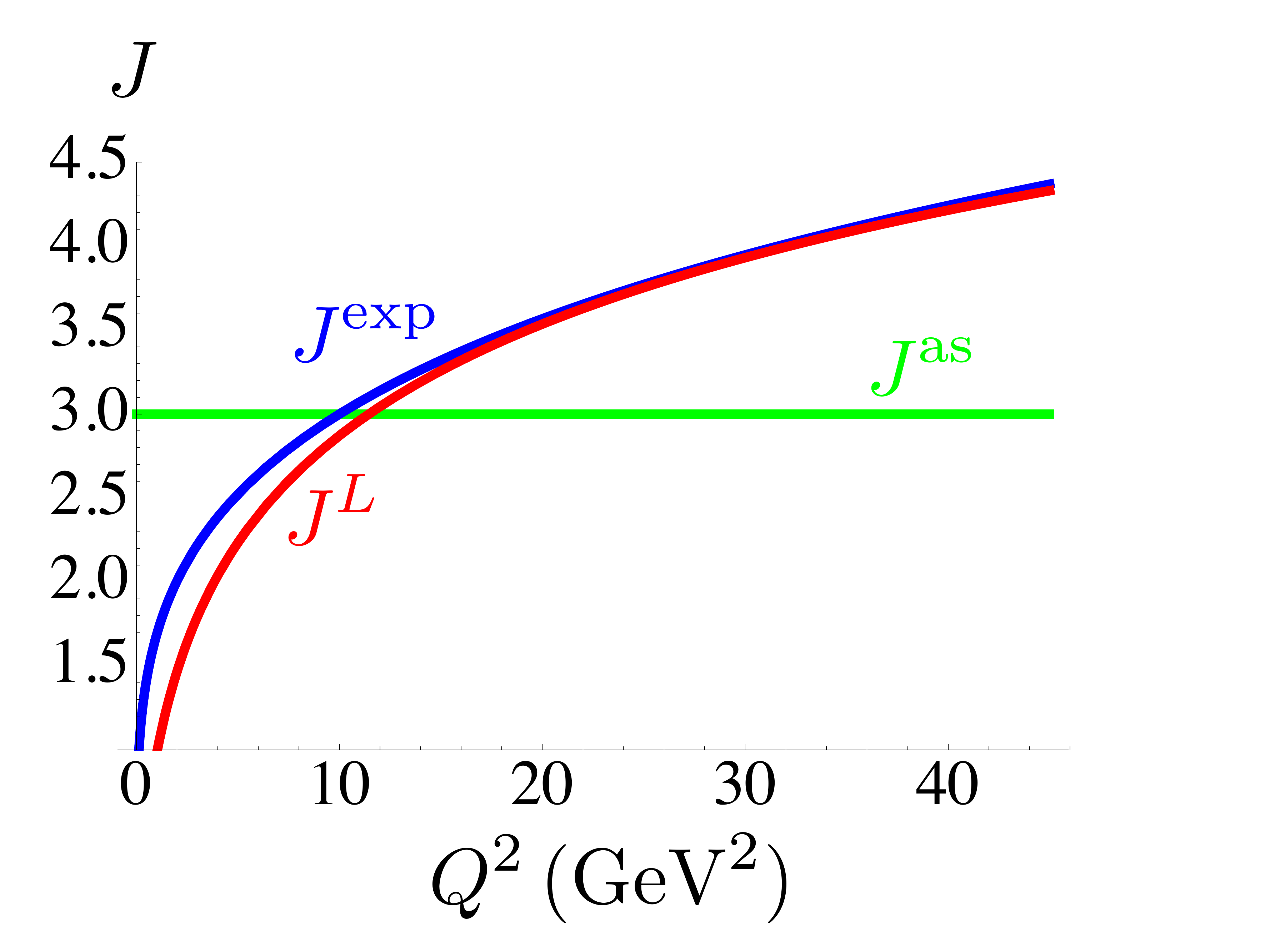}}
\caption{Left: {\sc BaBar} data. Right: Fit  $J^{\rm exp}(Q^2)$ shown with asymptotic 
prediction $J^{\rm as} =3$ and $J^L(Q^2)$.}
\label{babar}
\end{figure}
Unexpectedly, {\sc BaBar}   data   \cite{Aubert:2009mc}   
 are well  described by a logarithmic function \cite{Radyushkin:2009zg}
\begin{align}
 J^{L} (Q^2) = \ln \left ( \frac{Q^2}{M^2}+1 \right ) =  Q^2  \int_0^1 \frac{dx}{xQ^2+M^2} \  ,
\end{align}
 if one takes $M^2=0.6$\,GeV$^2$.
Notice  that $ J^{L} (Q^2) $   can be   obtained 
if one uses a  flat DA $\varphi_\pi (x) =f_\pi$ and 
changes \mbox{$xQ^2 \to xQ^2+M^2$}  in the pQCD expression  (\ref{8b}), 
expecting  that $M^2$  is generated by 
 transverse momentum. 
 However,   the $1/xQ^2 \to 1/(xQ^2 + M^2)$ modification 
is bringing in, {\it before the integration over $x$,}  a tower of 
 higher twist $(M^2/xQ^2)^n$ power corrections.
But it is known \cite{Musatov:1997pu} that the handbag diagram, 
because of its simple singularity structure near the coordinate  light cone  $z^2=0$, cannot generate an  infinite 
tower of power corrections: there are just two  power terms in the OPE for the handbag contribution, those 
generated by twist-2  and twist-4 operators.

\section{TMDA expression for  handbag  amplitude}

To  correctly include transverse momentum effects,
we parametrize the relevant matrix element
(taking scalar case to simplify   notations, for spin-1/2 quarks formulas are the same unless stated) 
\begin{align}
 \langle p |   \phi(0) \phi (z)|0 \rangle  \equiv F( (pz),z^2)
= & 
\int_{0}^{\infty} d \sigma \int_{0}^1 dx\, 
 \Phi (x,\sigma) \,  \,  e^{i  x (pz) -i \sigma {(z^2-i \epsilon )}/{4}}  \label{phisig}
 \end{align} 
through  the  {\it virtuality distribution amplitude}  (VDA)  $\Phi (x,\sigma)$. After that,  the 
coordinate 
$d^4z$  integral (see Fig. \ref{pigg}) can be taken, and we get   the 
handbag diagram contribution  in the VDA representation
 \begin{align}
T(Q^2) = &
 \int_{0}^1 \frac{dx}{xQ^2} \, \int_{0}^{\infty}   d \sigma \, 
  { \Phi (x,\sigma)    } 
\left \{ 1- e^{-[ixQ^2 + \epsilon]/  \sigma }  \right \}  \  . 
\end{align} 
The first term here corresponds to the twist-2 approximation, with  the $\sigma$-integral of 
 $
  { \Phi (x,\sigma)    } $  giving the pion DA 
$\varphi_\pi (x)  
$. 
The second term combines the contributions of ``invisible'' higher-twist  operators
uncapable to produce power corrections.  However,  after integration over $\sigma$ this term 
produces  a nontrivial function of 
$xQ^2$.  To interpret this result in terms of transverse momentum,  we   first specify   that
the  pion momentum $p$  has no transverse components  becoming pure  $p^+$
for $p^2=0$, and then project the matrix element on 
$z^+ =0$ plane by   taking  $z= (z^-, z_\perp)$
  \begin{align}
 \langle p |   \phi(0) \phi (z)|0 \rangle  |_{z^+=0, p_\perp =0} 
   = 
 \int_{0}^1 dx \,  {\varphi } (x, z_\perp)
 \, e^{i x (pz^-) }  \  . 
 \label{ida}
\end{align} 
This defines the 
  {\it impact parameter distribution amplitude}  (IDA) 
$  {\varphi } (x, z_\perp) $ whose  $z_\perp$ Fourier transform 
gives the {\it transverse momentum dependent
distribution  amplitude} (TMDA)  $ {\Psi}(x, k_\perp )$
\begin{align}
{\varphi } (x, z_\perp) = \int  {\Psi}(x, k_\perp ) \, e^{i (k_\perp z_\perp)}
\,  {d^2 k_\perp } =\int_{0}^{\infty} d \sigma  \,
 \Phi (x,\sigma) \,  \,  e^{i  \sigma {(z_\perp^2+i \epsilon )}/{4}} \  . 
\end{align}
  The crucial fact is that  TMDA can  be  written     in terms of 
  the covariantly defined VDA  
\begin{align}
{\Psi} (x, k_\perp ) =&   \frac{i }{\pi }
\int_{0}^{\infty} \frac{d \sigma }{\sigma} \, 
 \Phi (x,\sigma) \,  \,  
 e^{- i (k_\perp ^2-i \epsilon )/ \sigma} 
 \ . 
\end{align} 
As a result, the handbag term may be  written in terms   of TMDA. For spin-1/2 quarks, it reads 
 \begin{align}
 T(Q^2) 
= &  
   \int_{0}^1   \frac{dx}{xQ^2}   \int_{k_\perp^2 \leq {x} Q^2}    \Psi (x, { k}_\perp   ) \left [ 1- \frac{k_\perp^2}{xQ^2} \right ]
 \, d^2 { k}_\perp 
\ . 
 \end{align} 
One can see here the twist-2 and twist-4 terms as explicit  power-like contributions.
The hidden higher twist contributions correspond  then 
 to the (minus) integral  over the region $k_\perp^2 \geq xQ^2$. Now,  
 if  $\Psi (x, { k}_\perp   )$ decreases faster than any power of 
 $1/k_\perp^2$ for large $k_\perp$ (a usual   assumption for the 
 nonperturbative part of $\Psi (x, { k}_\perp   )$),  then the corrections
 to the twist-2 and twist-4  term {\it inside} the integral over $x$ would decrease 
 faster than any power of $1/Q^2$ for large $Q^2$,
 i.e. they have the property expected from ``invisible'' 
 contributions.

\section{Modeling transition form factor}

Generic VDA representation  (\ref{phisig})  treats $(pz)$ and $z^2$ as independent variables,
so we have no  
 {\it a priori} reasons for a particular 
 correlation of $x$ and $\sigma$ dependence in VDA. 
Thus, we try the simplest example: factorized models  for   VDAs, which result in   factorized models  for   TMDAs, 
$
\Phi (x, \sigma) =\varphi (x) \,  \Phi (\sigma)
\Rightarrow 
\Psi (x, k_\perp) = \varphi (x) \, \psi (k_\perp^2)/\pi  
$
in which 
the  $x$-dependence and  $k_\perp$-dependence appear in  separate factors.
 Assuming a Gaussian dependence on
$k_\perp$  for the TMDA results in a Gaussian dependence on $z_\perp$ for the  IDA 
  \begin{align}
\Psi_G (x, k_\perp) = \frac{\varphi (x)}{\pi \Lambda^2}  e^{-k_\perp^2/\Lambda^2}
\Rightarrow  \varphi_G (x, z_\perp) ={\varphi (x)} \, e^{-z_\perp^2 \Lambda^2/4}  \  . 
\label{gauss}
\end{align} 
 One may argue that a Gaussian fall-off for large $z_\perp$ 
is too fast   compared to  exponential  $\sim e^{-|z_\perp | m}$  behavior of  a 
 massive 
 propagator.  Some models with an exponential   fall-off were given in 
 Refs. \cite{Radyushkin:2014vla,Radyushkin:2014xaa,Radyushkin:2015wqa,Radyushkin:2015}.
 Here we will consider an 
 extreme $m=0 $ limit of these models, when 
\begin{align} 
\Phi_{m=0} (x, \sigma;\Lambda) = {\varphi (x)} \,  \frac{e^{i \sigma /\Lambda^2  
  -\epsilon \sigma} }{ i \Lambda^2 }  \  \  \   {\rm giving} \  \  \  
{\varphi }_{m=0} (x, z_\perp) =
\, \frac{\varphi (x)}{1+ z_\perp^2   \Lambda^2/4 }   \  . 
  \end{align}
Note that the  $z_\perp^2$ term of the 
$z_\perp$ expansion  of  ${\varphi } (x, z_\perp)$
in this model was adjusted to coincide with that of the exponential model,
so that    $\Lambda^2$ has the same 
meaning of  the  scale of $\bar \psi D^2 \psi$  operator.

We can use   these models now to calculate  the higher twist
contributions  to the transition form factor. 
In the case of  the Gaussian model (\ref{gauss}), we have (for spin-1/2 quarks) 
 \begin{align}
F_G (Q^2) = &
 \int_{0}^1 \frac{dx}{xQ^2}   \, 
{\varphi (x)} \, 
 \left [ 1-  \frac{\Lambda^2}{xQ^2}  \left (1 -   e^{-xQ^2/\Lambda^2} \right ) \right ]   \  .
\label{FGauss2}
\end{align} 
Again, for  large $Q^2$,  Eq. (\ref{FGauss2}) 
 displays  a  power-like twist-4 correction and the 
 term that  corresponds  to the region  $k_\perp^2 \geq xQ^2$ 
accumulating  
contributions of ``invisible''   operators with twist 6 
and higher. 
Note that,   despite of the  $1/x^2$
singular term in the integrand, the integral (\ref{FGauss2})  is finite 
for $\varphi (x)$ as singular as $1/x^{1-\alpha}$ with  $\alpha>0$,
which includes a flat DA (with $\alpha=1$).
Furthermore, 
 a formal $Q^2 \to 0$ limit is finite: 
$
F_G (Q^2=  0) 
= { f_\pi }/ { 2 \Lambda^2}  
$.   In fact, $F(Q^2)$ is finite for $Q^2=0$ in any model with finite
$\Psi  (x, 
k_\perp=0)$.  For the non-Gaussian $m=0$ model, we have 
 \begin{align}
F (Q^2) = &
 \int_{0}^1  \frac{dx}{xQ^2} \, 
{\varphi (x)} \, 
\left [ 1-    \frac{\Lambda^2}{xQ^2}  + 2  
K_2 (2 \sqrt{x }Q/\Lambda) 
 \right ] 
 \  ,  
\end{align} 
where $\Lambda^2$ is again the scale characterizing the $\bar q D^2 q$ 
matrix element.    

 To describe the  {\sc BaBar} data we use a  flat DA $\varphi  (x)  = f_\pi$ 
 with the    scale $\Lambda_G^2=0.35\,$GeV$^2$ for the Gaussian model and 
 $\Lambda_{m=0}^2=0.6\,$GeV$^2$ for the $m=0$ non-Gaussian model. 
Fitting the  Belle data requires a more narrow DA  $\varphi  (x)  \sim  f_\pi (x\bar x)^{0.4}$
 with the scale $\Lambda_G^2=0.3\,$GeV$^2$ 
   for the Gaussian model
 and  a larger value $\Lambda_{m=0}^2=0.4\,$GeV$^2$ 
  for the $m=0$ non-Gaussian model. 
  
\begin{figure}   
\centerline{\includegraphics[width=2in]{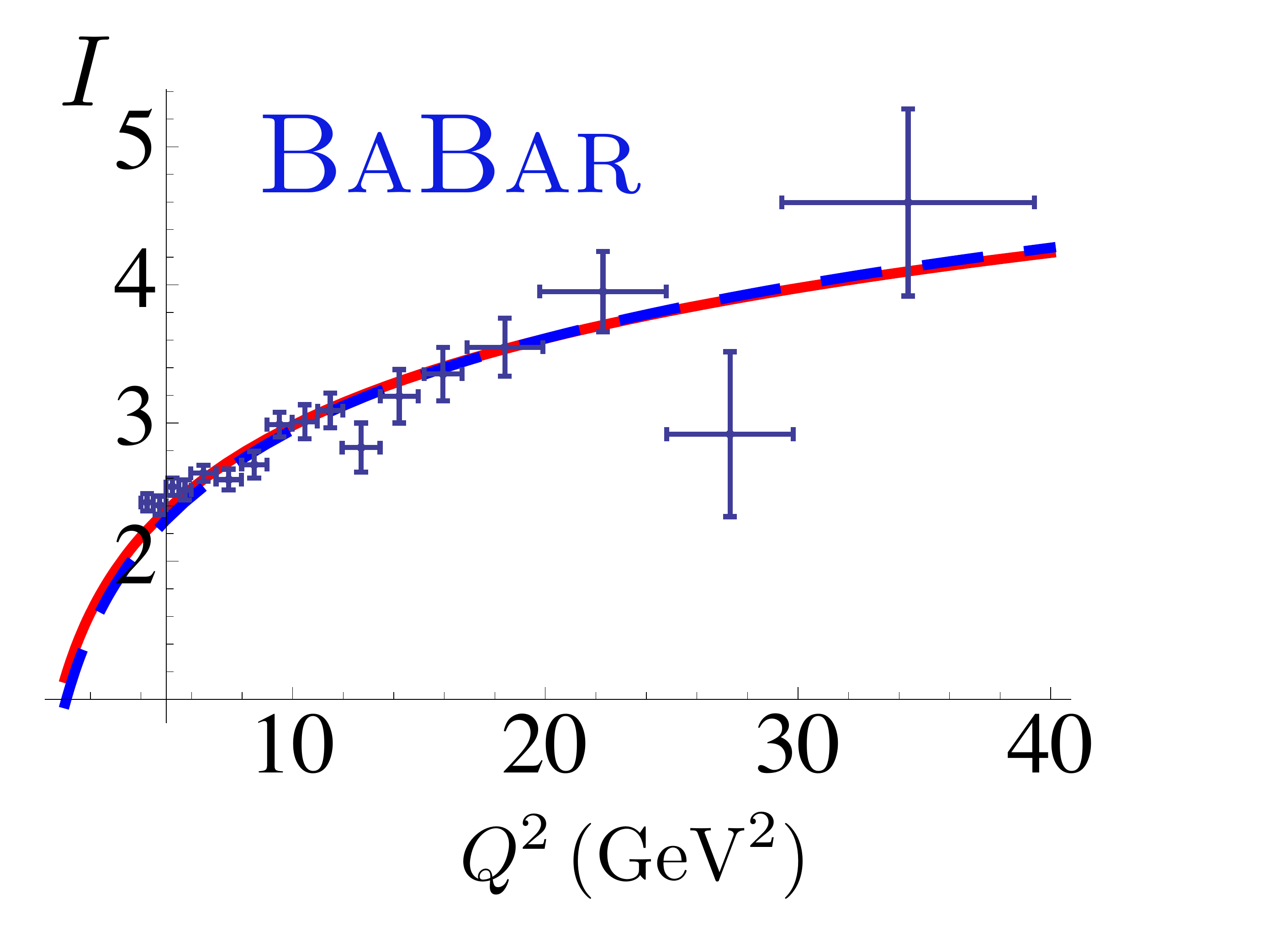} \hspace{1cm}  \includegraphics[width=2in]{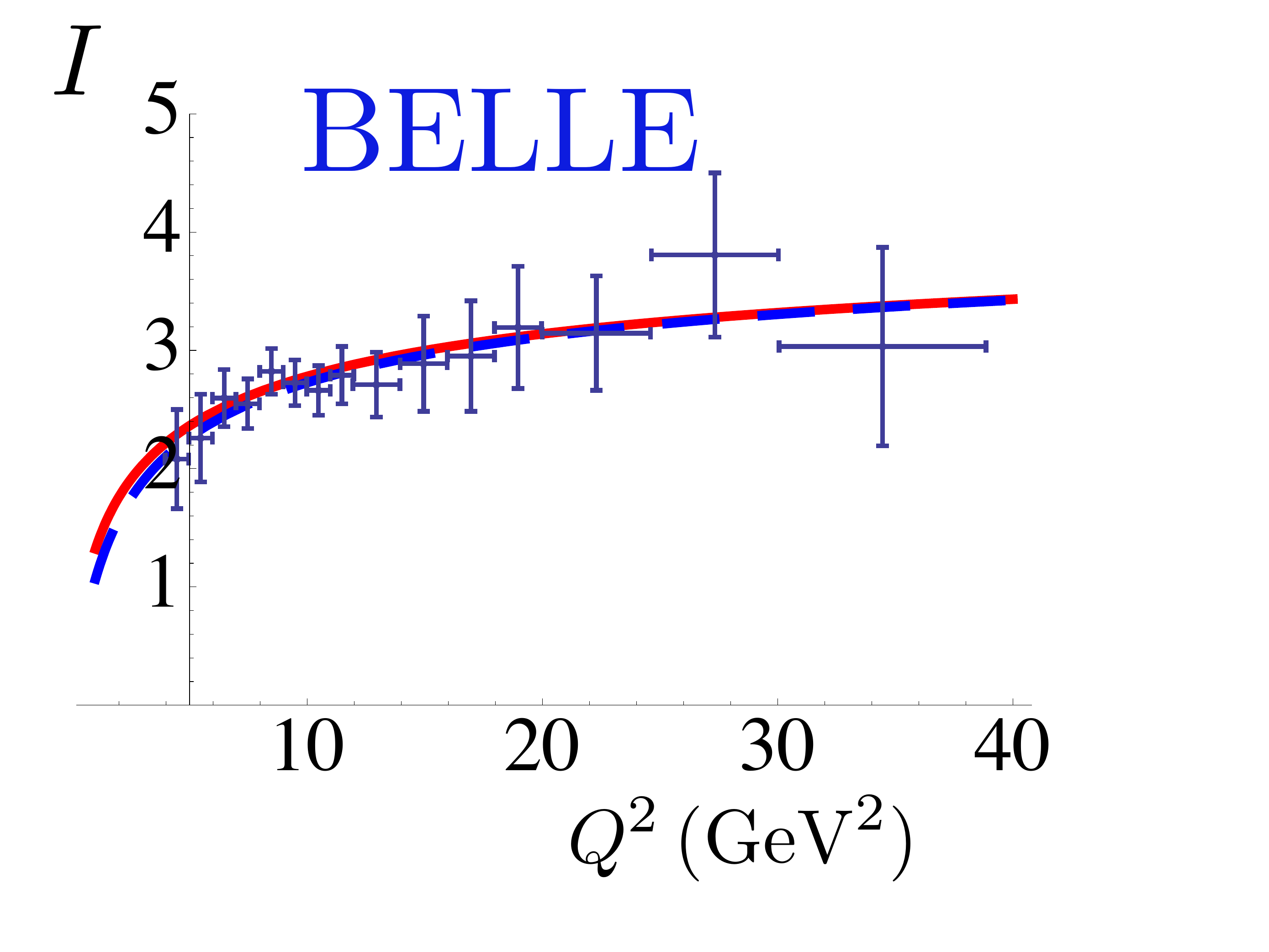}}
\caption{{\sc BaBar} and Belle  data compared to model curves  described in the text.}
\label{ggpidata} 
   \end {figure}

  A close agreement of our models with the data  (see Fig. \ref{ggpidata}) 
  for a very wide range of $Q^2$ indicates that the ``invisible contributions''
  correctly describe the physics behind the nontrivial shape 
  of the experimental curves for $Q^2 F(Q^2)$.   Note also that one has 
   $J=4.5$ for the DA $\varphi  (x)  \sim  f_\pi (x\bar x)^{0.4}$ used to describe 
 the   Belle data, which means that the 
curve $J(Q^2)$ is   well below its  asymptotic value even 
for $Q^2 \sim 40$ GeV$^2$.  Within the standard 
pQCD picture, in which the form factor $F(Q^2)$  is 
a sum of  a few power-like terms, and 
given the size  $\Lambda_G^2=0.3\,$GeV$^2$  or $\Lambda_{m=0}^2=0.4\,$GeV$^2$ 
of the scale involved, it is simply  impossible to understand 
such a slow approach to the asymptotic value.

 Another point is that in the region 
  $Q^2 \lesssim 10$ GeV$^2$, where the BaBar and Belle data agree,  the data 
may be described by models with both flat and $\sim  (x\bar x)^{0.4}$
behavior.  In other words, the data alone cannot tell us what is the shape of the pion DA,
even when the data are in a region of $Q^2$  by  two  orders of magnitude larger
than $\Lambda_{\rm QCD}^2$.

  \section{One-gluon-exchange diagram for the pion  form factor}

\begin{figure}   
  \centerline{\includegraphics[width=2in]{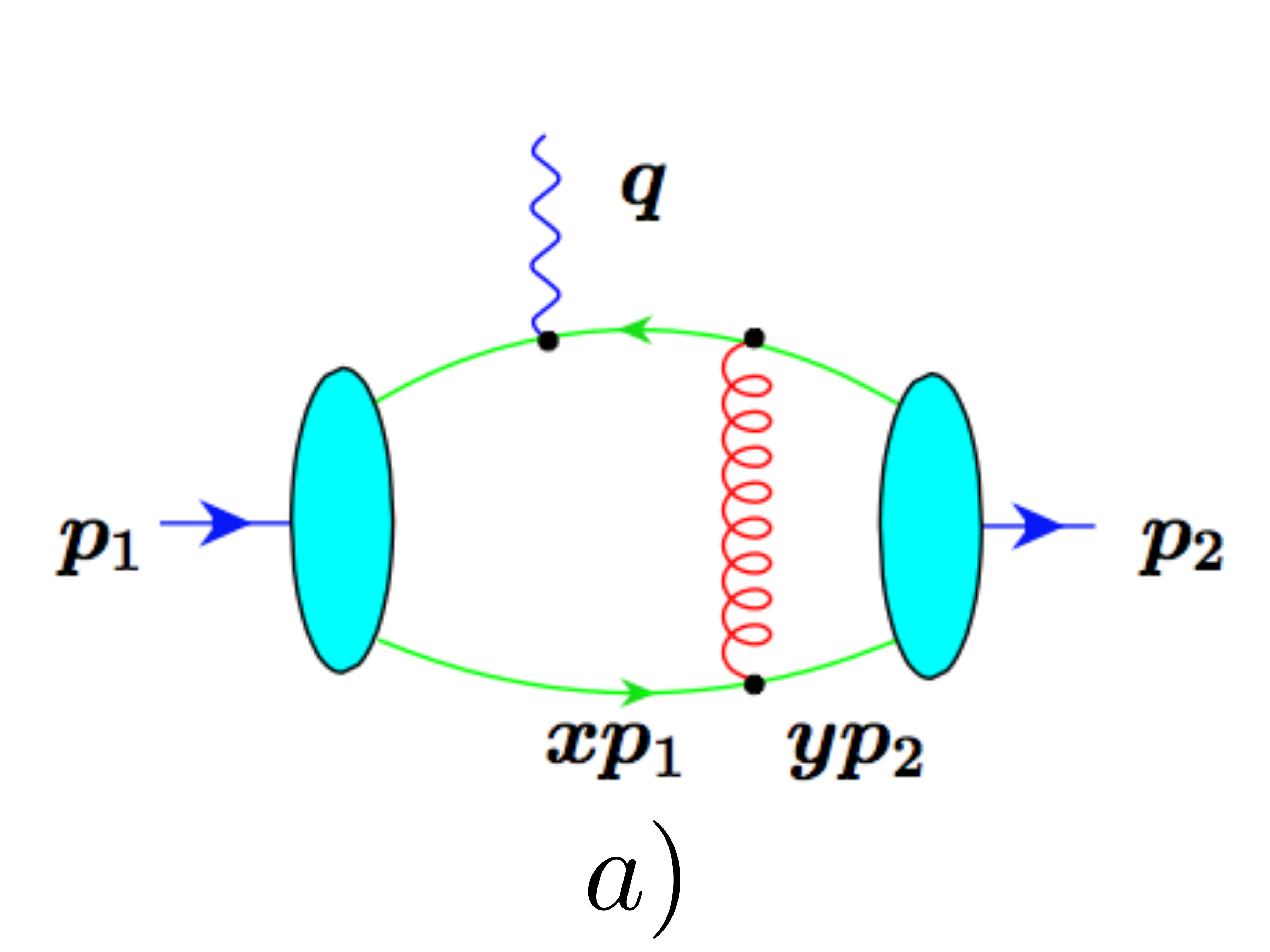}  \   \  \includegraphics[width=1.8in]{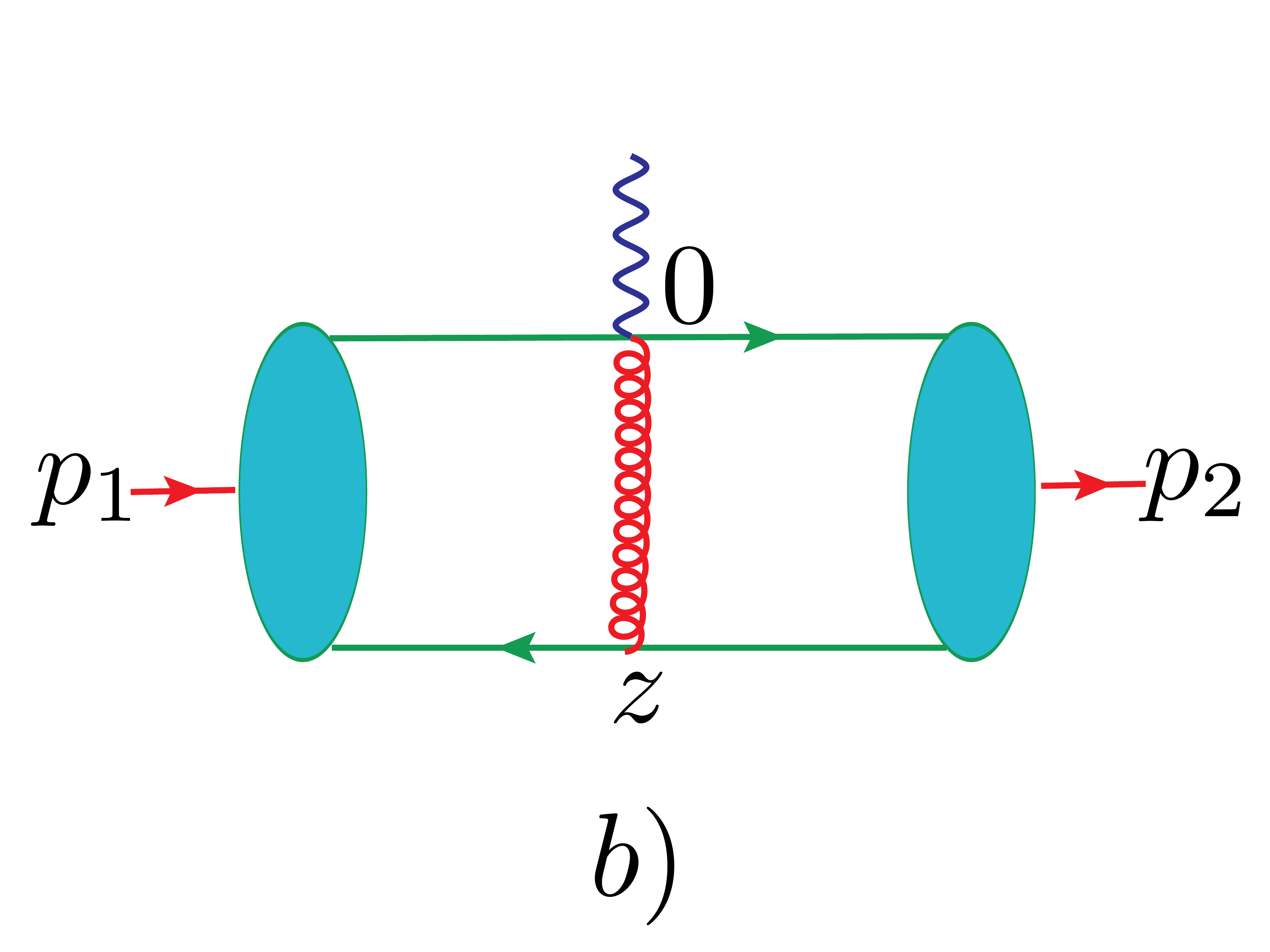}}
\caption{  Perturbative QCD  one-gluon-exchange contribution to the  pion EM form factor} 
\label{pionemff} 
   \end {figure}

Let us now apply  the VDA formalism to the   
 hard   contribution for  the pion electromagnetic (EM) form factor. 
  The short-distance subprocess in this case contains one quark and one gluon 
  propagator, with virtualities $xQ^2$ and $xyQ^2$, respectively (see Fig. \ref{pionemff}a).
  
  In the twist-2 expression for spin-1/2 quarks, 
  the denominator factor $xQ^2$ of the hard quark propagator 
  is cancelled by the $xQ^2$ factor  coming from the numerator trace,
  so that the final  integral 
  \begin{align}
 F_\pi^{\rm as \, (pQCD)}(Q^2)  =& \frac{8 \pi \alpha_s  }{9} \int_0^1 dx \int_0^1 dy \
\frac{ \varphi_\pi  (x) \, \varphi_\pi  (y)}{xyQ^2}  = 2 \frac{s_0}{Q^2} 
\,\frac{\alpha_s}{\pi}  (J/3)^2
\nonumber 
\end{align}
 (where $s_0 = 4 \pi^2 f_\pi^2 \approx 0.7$ GeV$^2$)   is given by an expression 
  which may be thought of as coming from a diagram 
 where  the hard quark propagator is shrunk into a point (see Fig. \ref{pionemff}b).
 Again, the pQCD  result diverges for flat DA $\varphi_\pi  (x) =f_\pi$. Since the 
virtuality of  exchanged gluon is   $xyQ^2$, one may naively 
expect  $xyQ^2 \to xyQ^2 + 2 M^2$ 
when the transverse momentum is included. As we have learned,
in the VDA approach
such   effects are desribed by  formulas  more complicated  than that.

 To illustrate  the VDA modifications, we consider the 
  simplified diagram shown in  Fig. \ref{pionemff}b  that has the same  asymptotic 
  expression as the full pQCD diagram Fig. \ref{pionemff}a. 
It may be written as 
       \begin{align}
 T(p_1,p_2) =&a   \int_{0}^1 dx  \int_{0}^1 dy
 \int {d^4z}\, e^{-ix(p_1z)+i y (p_2z) }   D_c(z) \, 
  \, B(x, z^2/4) \, B(y, z^2/4) \ . 
 \end{align}
     in the coordinate representation (we denote $a= 8 \pi \alpha_s/9$).   In terms of  VDAs,  we obtain 
    \begin{align}
 F(Q^2) = &a 
 \int_{0}^{\infty}   d \sigma_1 \int_{0}^{\infty}   
 d \sigma_2  \int_{0}^1  \Phi (x,\sigma_1) \int_{0}^1   \Phi (y,\sigma_2)  
\frac{dxdy}{xyQ^2} \, 
 \left [ 1- e^{-[ixyQ^2 + \epsilon]/ ( \sigma_1+\sigma_2) }  \right ]\  , 
    \end{align}

  The first term in the square brackets does not depend on $\sigma_1, \sigma_2$ and 
 produces   the 
 twist-2 expression. Using  the definition (\ref{ida}) of IDA $\varphi (x,b_\perp)$ we arrive at 
  \begin{align}
 F(Q^2) = &\frac{a}{(2\pi)^2}  \int_{0}^1 dx \int_{0}^1{dy}  \, 
 \int_{k_\perp^2 \leq xyQ^2}  \frac{d^2 k_\perp} {xyQ^2}  \, 
 \int  e^{i (k_\perp b_\perp)} \, 
   \varphi (x,b_\perp)\,   \varphi (y, b_\perp)   \, 
   d^2 b_\perp\  . 
   \label{fpiIDA}
  \end{align}
  The $k_\perp^2 \leq xyQ^2$ restriction converts into the Bessel function
 $J_1$ in the  impact parameter space 
giving 
  \begin{align}
 F(Q^2) = & a\int_{0}^1 dx \int_{0}^1{dy}  \, 
 \int_{0}^\infty   \frac{  d b_\perp} {\sqrt{xy Q^2} }  \,   
 J_1 (b_\perp\sqrt{xyQ^2 })\,  \varphi (x,b_\perp)\,   \varphi (y, b_\perp)   \,    . 
 \label{FpiJ1}
  \end{align}
  Taking  the factorized Ansatz  with a Gaussian (G)  or power-law (P) dependence on $b_\perp^2$   \begin{align}
\varphi_G (x, b_\perp) ={\varphi (x)} e^{-b_\perp^2 \Lambda_G^2/4} \ \ , \ \ 
\varphi_P (x, b_\perp) ={\varphi (x)} / (1+ b_\perp^2 \Lambda_P^2/4)
\end{align} 
we obtain the following expressions  
  \begin{align}
 F_G(Q^2) = &  {a} 
   \int_{0}^1 \frac{dx \, dy} {xyQ^2} {\varphi (x)}  
 {\varphi (y)}  \,    
     \left [ 1- e^{-xy Q^2/2 \Lambda_G^2} \right ] \equiv a\frac{ f_\pi^2}{Q^2}  I_G (Q^2) = 2 \frac{s_0}{Q^2} 
 \frac{\alpha_s}{\pi} \,     \frac{I_G (Q^2)}{9} \  , 
  \\
     F_P(Q^2) = &  {a} 
   \int_{0}^1 \frac{dx \, dy} {xyQ^2} {\varphi (x)}  
 {\varphi (y)}  \,  
     \left [ 1- 2 \, \frac{xy Q^2}{ \Lambda_P^2} \, K_2 ( 2 \sqrt{xy} Q/\Lambda_P)  \right ] 
     \equiv a\frac{ f_\pi^2}{Q^2}  I_P (Q^2)
  \   . 
  \end{align} 
Just like in the transition form factor case, these integrals  are finite for $Q^2=0$, with   the values  
\mbox{$
 F_G(Q^2=0) =  a\,  {f_\pi^2 }/{2 \Lambda_G^2}
 $,}   $F_P(Q^2=0) =  a\,  {f_\pi^2 }/{ \Lambda_P^2}$ . 
 
\begin{figure} 
  \centerline{\includegraphics[width=2in]{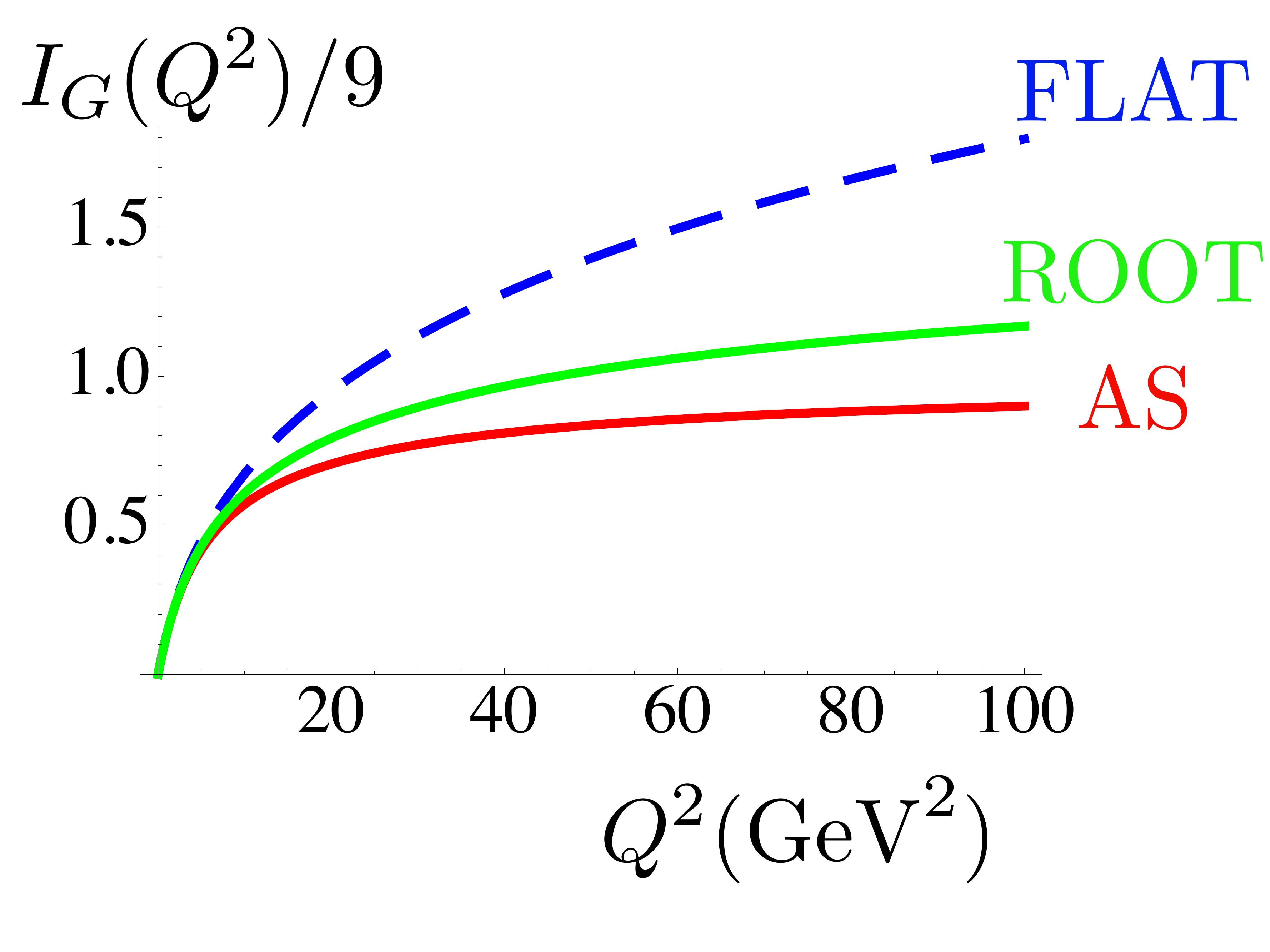} \hspace{1cm}  \includegraphics[width=2in]{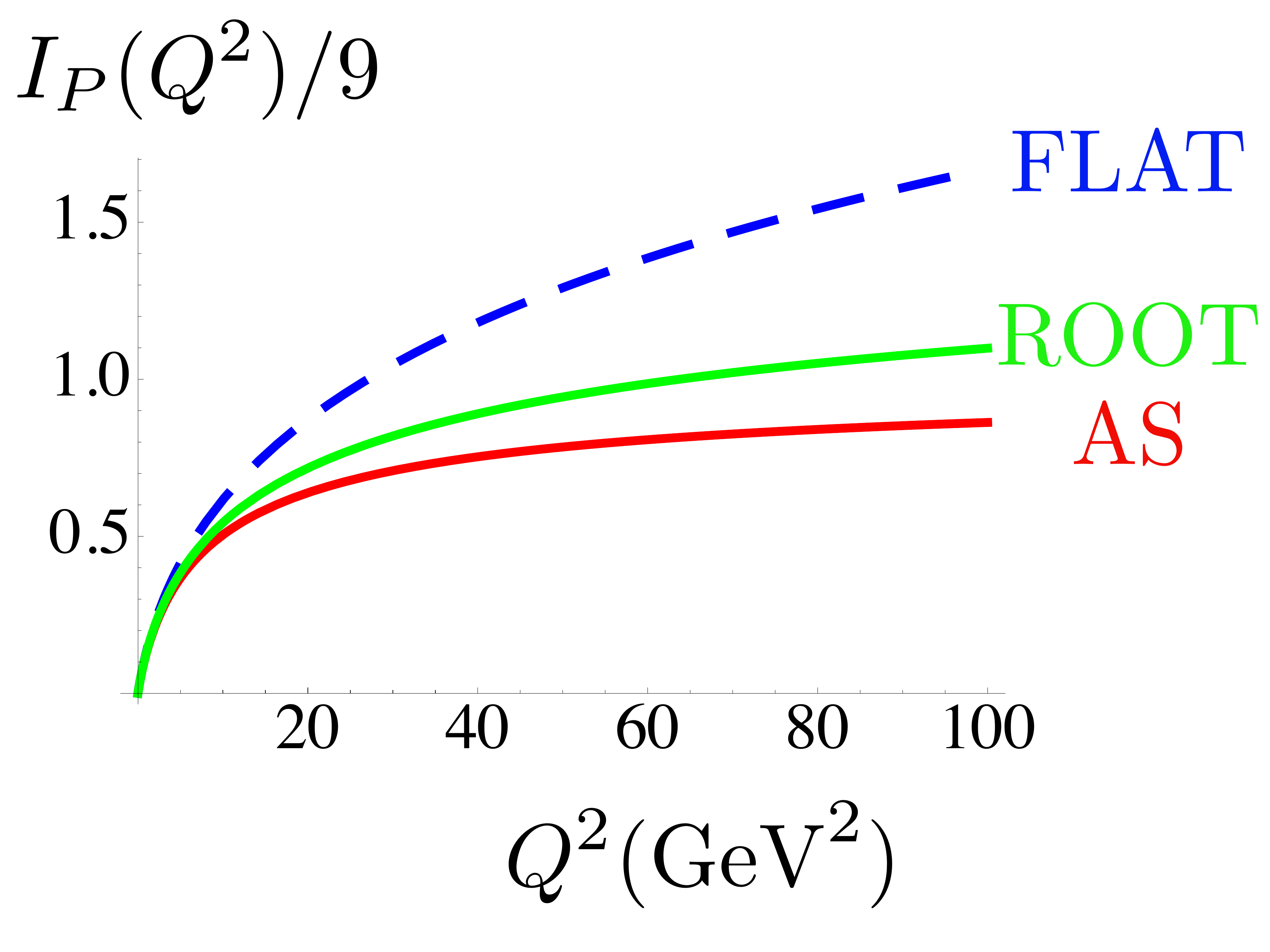}}
   \caption{Functions $I(Q^2)$ for Gaussian (left) and power-like (right) IDA models for $Q^2$ up to 100 GeV$^2$.}
    \label{em100} 
   \end {figure}

\begin{figure} 
 \centerline{\includegraphics[width=2in]{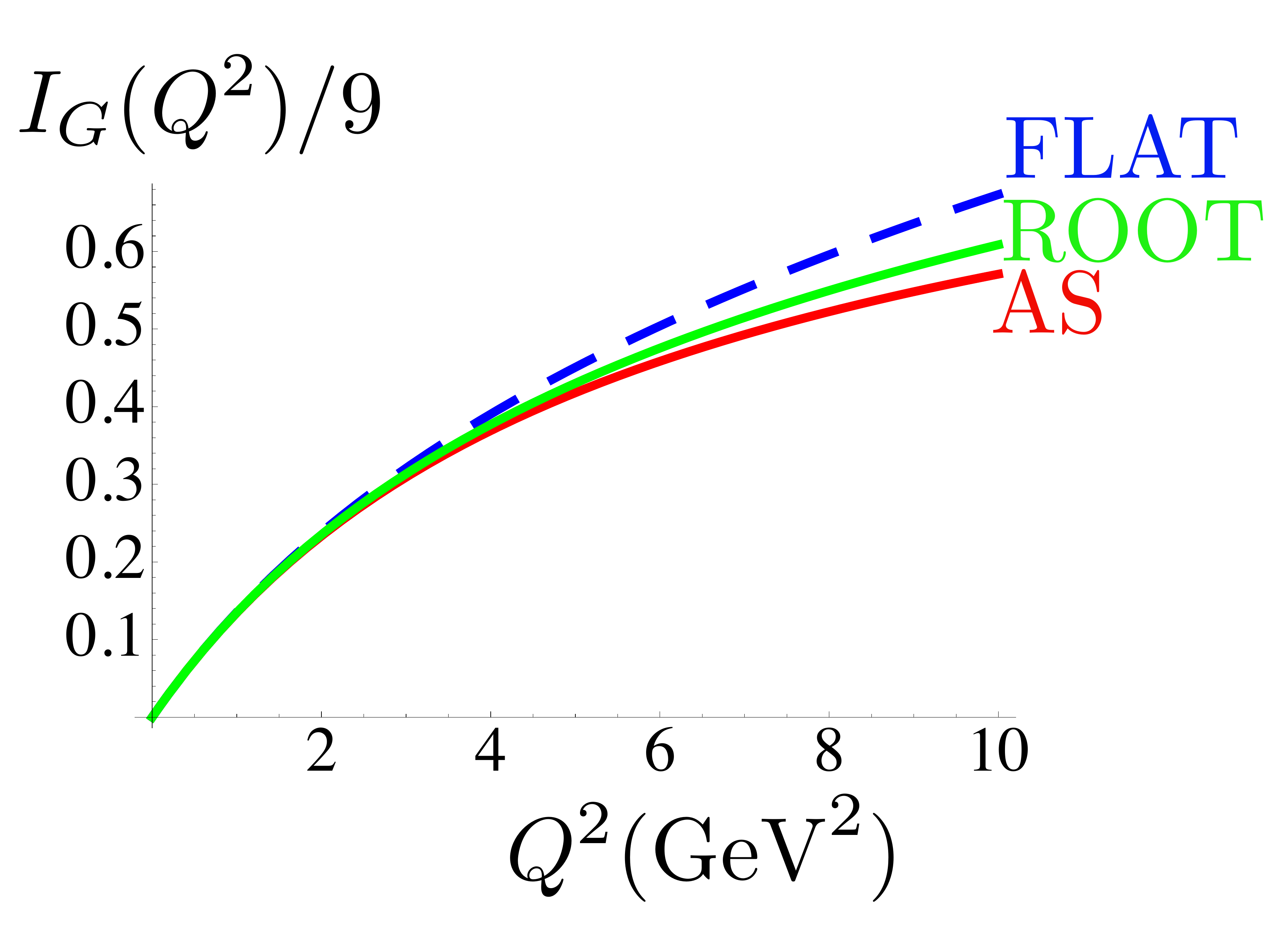} \hspace{1cm}  \includegraphics[width=2in]{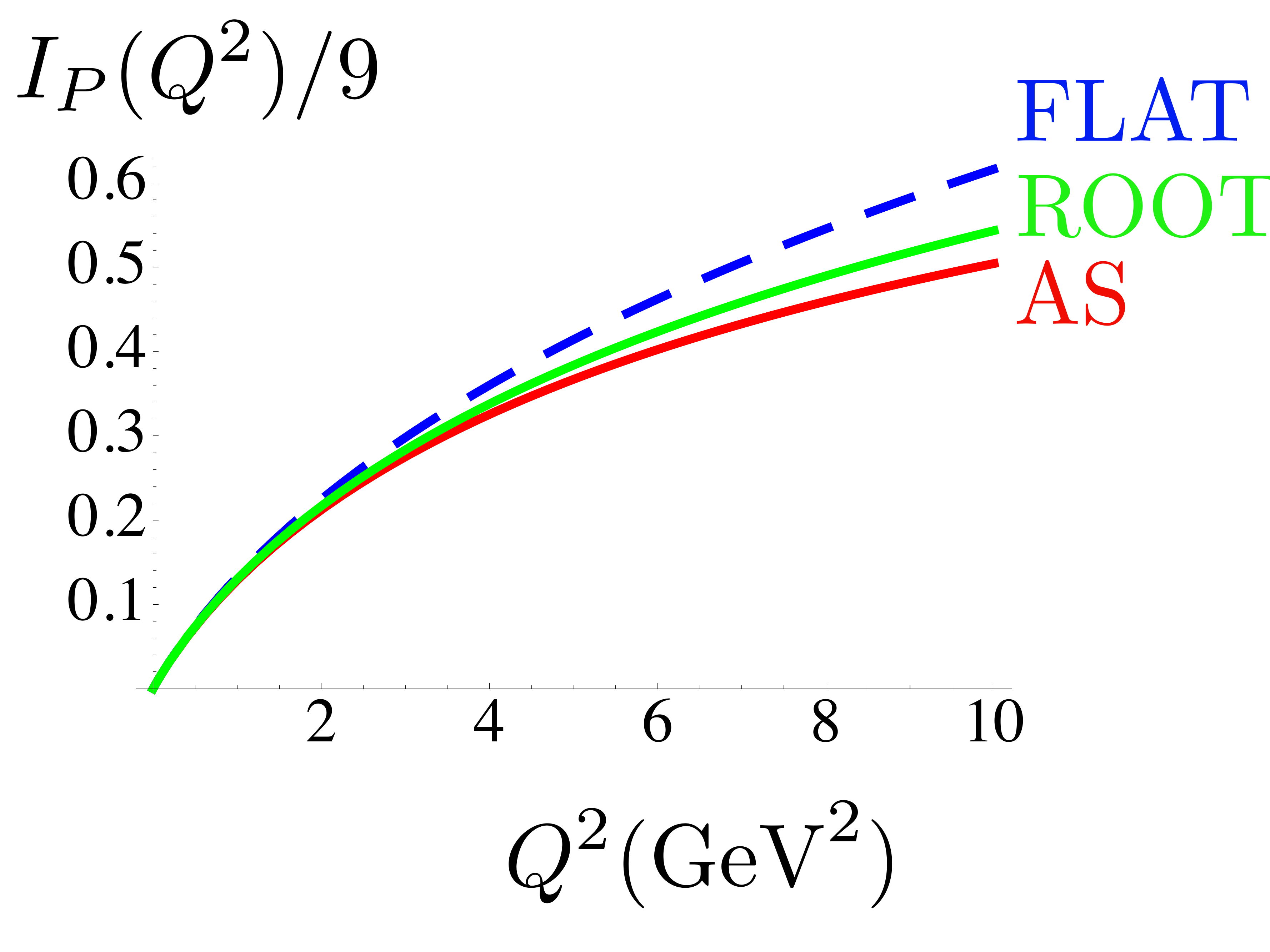}}
   \caption{Functions $I(Q^2)$ for Gaussian (left) and power-like (right) IDA models  for $Q^2$ up to 10 GeV$^2$.}
    \label{em10} 
   \end {figure}

 Taking for definitness the scales  $\Lambda_G^2 =0.35$ GeV$^2$ and \mbox{$\Lambda_P^2 =0.6$ GeV$^2$}  
 of the same size as those 
 that were fitting the BaBar transition form factor data, 
we  plot the ratio $I (Q^2)/9$  for both cases,  see Figs. \ref{em100}, \ref{em10}.
 To show the  sensitivity to the shape of the pion DA, we  use  3 different choices for 
$\varphi (x)$, namely, flat, ``root'', and asymptotic.  
 Notice that 
 ``1''  on these plots  would correspond  to   pQCD prediction with asymptotic DA. 
 One can see from Fig. \ref{em100} that it requires pretty large $Q^2\gtrsim 20$ GeV$^2$ 
 to clearly discriminate between the 3 curves.  To emphasize this point, we repeat the plots 
 on Fig. \ref{em10}, 
 restricting the range of $Q^2$ to experimentally accessible values of  $Q^2 \lesssim 10$ GeV$^2$. 
 One can see that the 3 curves are practically indistinguishable in this region. 
 To understand this outcome, we note  that the curves are still rising with $Q^2$ and are 
 almost linear. Thus, we conclude that the average gluon virtuality 
 $ \langle xy Q^2 \rangle$ in this region is practically constant, e.g. in Gaussian case
 $ \langle 1/(xy Q^2 )\rangle =  [2\Lambda_G^2 
 + \langle x \rangle  \langle y  \rangle Q^2/2  +\ldots]^{-1}  = [2\Lambda_G^2  + Q^2/8+\ldots]^{-1}$ . 
 It starts to become visibly proportional to $Q^2$ only when the curves on Fig. \ref{em100} 
 flatten, i.e. well above $Q^2 \sim 10$ GeV$^2$.  
 
 In other words,  we see no chances 
 to detect  experimentally the $\sim 1/Q^2$ perturbative QCD  prediction for the ``hard'' gluon-exchange 
 contribution.  Furthermore, we observe  that the shape of the pion DA is not important as far as 
  $Q^2 \lesssim 10$ GeV$^2$: the result for the flat  DA in this  region  
 is practically the same as for the asymptotic DA.  Moreover, because the
 asymptotic behavior sets very slowly,  the absolute  value of $I(Q^2)/9$ 
 is well below  ``1''  corresponding to the formal pQCD prediction, approaching 
 just a half of it for the highest $Q^2 \sim 10$ GeV$^2$   reachable at Jefferson Lab.

Numerically,  using $2s_0\approx 1.3$ GeV$^2$   and assuming a ``frozen''  value $\alpha_s/\pi =0.1$,
we have  $Q^2 F_\pi^{\rm gluon \ exchange} (Q^2) = 0.13 \, [I(Q^2)/9]$ GeV$^2$.
Thus, our estimate gives $Q^2 F_\pi^{\rm gl. \ ex.} (Q^2) \lesssim  0.04$ GeV$^2$ for $Q^2=2.45$  GeV$^2$
 or more than 10  times 
below the  existing JLab measurement \cite{Horn:2006tm}.  Though  this outcome doubles  to 
 0.08 GeV$^2$ for $Q^2=10$  GeV$^2$, it  still does not look  significant. 
 
One may ask:  if the  gluon exchange contribution is so small, 
what is the mechanism explaining observed 
pion EM form factor  behavior?  The answer was given long time ago \cite{Nesterenko:1982gc}: it is the Feynman mechanism, 
corresponding (in the light-front terminology)  to the Drell-Yan \cite{Drell:1969km}  overlap of soft wave functions.
However,  this is a subject of a different talk and, hopefully,  another direction for applications of the VDA formalism!


\section*{Acknowledgements}

This work is supported by Jefferson Science Associates,
 LLC under  U.S. DOE Contract No. DE-AC05-06OR23177
  and by U.S. DOE Grant \#DE-FG02-97ER41028.

\bibliographystyle{JHEP}

\bibliography{pionem.bib}

 \end{document}